# Optically enriched and guided dynamics of active skyrmions

HAYLEY R. O. SOHN,[1] CHANGDA D. LIU,[1] ROBERT VOINESCU,[1] ZEZHANG CHEN,[1] AND IVAN I. SMALYUKH[1,2,3*]

[1]*Department of Physics and Materials Science and Engineering Program, University of Colorado, Boulder, CO 80309, USA*
[2]*Department of Electrical, Computer, and Energy Engineering and Soft Materials Research Center, University of Colorado, Boulder, CO 80309, USA*
[3]*Renewable and Sustainable Energy Institute, National Renewable Energy Laboratory and University of Colorado, Boulder, CO 80309, USA*
*[*ivan.smalyukh@colorado.edu](mailto:ivan.smalyukh@colorado.edu)*

**Abstract:** Light provides a powerful means of controlling physical behavior of materials but is rarely used to power and guide active matter systems. We demonstrate optical control of liquid crystalline topological solitons dubbed "skyrmions", which recently emerged as highly reconfigurable inanimate active particles capable of exhibiting emergent collective behaviors like schooling. Because of a chiral nematic liquid crystal's natural tendency to twist and its facile response to electric fields and light, it serves as a testbed for dynamic control of skyrmions and other active particles. Using ambient-intensity unstructured light, we demonstrate large-scale multifaceted reconfigurations and unjamming of collective skyrmion motions powered by oscillating electric fields and guided by optically-induced obstacles and patterned illumination.

## 1. Introduction

Active matter systems attract a great deal of recent interest because they provide the means of understanding many out-of-equilibrium phenomena in nature, enable new breeds of dynamic materials and promise to revolutionize technologies ranging from optofluidic to information displays [1-5]. The inanimate active matter systems range from colloids [2] to robots [3], lyotropic liquid crystals (LCs) [4,5], granular systems [6,7] and particle-like topological structures in physical fields with either singular [8] or solitonic continuous [9] configurations. While powered through conversion of chemical, mechanical or electrical energy into motions at the single particle level, these systems exhibit emergent collective behaviors that arise from many-body out-of-equilibrium interactions. By harnessing complex interactions of such systems with light, new means of light-matter interaction could be potentially envisaged. However, differently from conventional photo-responsive material systems, the means of guiding out of equilibrium phenomena in active matter remain limited, even though a few elegant demonstrations of such optical guiding of active matter already exist [10-14]. Optical guiding of active behavior in LCs could be of particular interest as it could allow for the ensuing control of light by these anisotropic materials, as is done in displays and spatial light modulators [15,16], while also invoking emergent many-body effects, such as schooling of skyrmions [9]. However, so far only equilibrium free energy landscapes for controlling behavior of these topological solitons have been tuned with light [17].

Topological solitons are localized field configurations that arise in nonlinear field theories, such as those found in particle physics and cosmology. Until recently, their realization and stabilization in experiments remained elusive [18,19]. Condensed matter systems have emerged as a favorable testbed for stabilization of topological solitons because of the facile ability to control the energetic landscape within materials such as solid-state magnets and LCs [20-32]. Chiral nematic LCs provide a favorable host for stabilization of twisted solitons, such as so-called skyrmions [28] and hopfions [20,21], because of the material's natural tendency to twist and the ensuing nonlinear nature of the free energy functional. [33-35] The simplest two-dimensional elementary skyrmion is an axi-symmetric translationally-invariant structure in which the field vectors exhibit π-twist radially outward from the center to the periphery [28]. Like their higher-dimensional counterparts, the two-dimensional elementary skyrmion cannot be eliminated or destroyed by smooth perturbations of the field and is therefore stable under a variety of conditions and external field manipulations [34]. One can drive skyrmion motion by electric currents in magnetic materials [26,27]. Skyrmion motion can also be realized in chiral LCs by means of rotational dynamics of the LC director field **n(r)** prompted by modulation of electric fields, in which the electric power is applied via electrodes on confining substrates, similar to those used in display technologies [9,15,16,36,37]. However, this electrically-induced active motion has only limited types of collective emergent behavior. Previous developments only allow for one "setting" of motion where all skyrmions tend to synchronize to move together at a constant velocity and in the same direction [9]. Inspired by complex dynamics that one observes in crowds of people and other active systems interacting with various obstacles and environments, we present an experimental approach for selective control of skyrmion motion using diverse optical manipulation techniques. Optical tweezers are used to create obstacles and to set directions in the field of motion whereas patterned blue light illumination, coupled with photo-tunable cholesteric pitch behavior [17], provides hands-off manipulation of the energetic landscape within the sample. The combination of these optical manipulation tools is used to guide, deflect, and reconfigure skyrmion motion. The electrically-powered skyrmion motion [36], optical manipulation using laser tweezers [34], and photo-manipulation using patterned light illumination [17,38] are combined to enhance collective motion controllability. We demonstrate reconfigurations of the dynamic skyrmion crowds into single-file lines, steer and inhibit skyrmion motion by changing polarization of patterned light illumination, optically induce skyrmion jamming and unjamming transitions, and other means of controlling complex motion with emergent active behavior. Since the system is based on materials and preparation techniques that are commonly used in the LC display industry [15,16], this work may lead to re-defining human-computer communications through LC-based touch-screen displays and expand the scope of augmented or virtual reality devices.

## 2. Materials and methods

We employ experimental implementation of samples reminiscent of LC displays, as is shown in the experimental schematic (Fig. 1a). These inch-square samples are created by infiltrating a chiral nematic LC mixture into a glass cell constructed by gluing two conductive substrates together and setting the cell gap with 10 μm glass spacers. The inner surfaces of the glass substrates have transparent conducting layers of indium-tin oxide and were pre-treated for finite-strength vertical surface boundary conditions via spin-coating with SE-1211 (purchased from Nissan) at 2700 rpm for 30 s. To induce crosslinking of the alignment layer, the substrates were then baked for 5 min at 90 °C and for 1 h at 190 °C. The LC mixture is composed of a nematic host with negative dielectric anisotropy, either MLC-6609 (Merck) or ZLI-2806 (EM Chemicals), that has been heated to the isotropic phase for thorough mixing with a chiral additive (Table 1). For experiments done in samples without photo-sensitivity to low-intensity blue light, the additive CB-15 from EM Chemicals was used. Alternatively, mixing the nematic hosts with the QL-76 additive (obtained from the Air Force Research Lab [39]) enabled photo-tuning of the chiral pitch, $p_0$, via blue light exposure. The ground state helicoidal pitch of these mixtures can be calculated using the relation $p_0 = 1/h_{HTP} \cdot c$ [40] (Table 1), where $h_{HTP}$ is the helical twisting power of the chiral additive in the host medium and $c$ is its concentration by weight. In order to stabilize skyrmions spontaneously, we set the pitch to be approximately equal to the cell gap of 10 µm. For the photo-responsive materials, the pitch can be increased from $p_0$ to excited-state value $p_e$ by blue-light illumination via a trans-cis isomerization that decreases the helical twisting power of the QL-76 dopant [41] which, in our samples, enables tunability from $p_0 = 10$ µm to $p_e \approx 22$ µm [17]. Because we use low intensity ~ 1nW per square micron [13] blue patterning light in the 450-480 nm range and thin ~ 10 µm LC films with chiral additive concentrations near 0.2-0.3 wt%, absorbance within the sample is negligible [41,42]. Therefore, we assume that the illumination has consistent non-diminishing intensity throughout the sample thickness.

We take advantage of the high stability of skyrmion structures in samples with cell thickness $d \approx p_0$, and use a number of techniques for spontaneous or systematic generation of these topological structures. For spontaneous generation, the samples were heated past each material's clearing temperature ($T_{NI}$, Table 1) to the isotropic phase and cooled rapidly using compressed air. Alternatively, skyrmions were generated by means of optical reorientation induced by a 1064 nm Ytterbium-doped fiber laser that comprises our "optical laser tweezer" setup (YLR-10-1064, IPG Photonics) [34-37,43]. This optical reorientation, known as the optical Fredericks transition [35], results in controlled generation of twisted structures when the laser beam is focused on the midplane of the sample and the power is tuned to > 50 mW, inducing local LC director realignment away from the homeotropic far-field background induced by the vertical surface boundary conditions. The same optical setup was also used to "pin" skyrmions to the cell substrates via altering the alignment layer when the focal plane of the laser focus is adjusted to be closer to one of the confining substrates, which in turn creates stationary solitonic obstacles within the sample. Using these techniques, skyrmions were selectively generated at ~50mW power and pinned at powers between 70 and 150 mW. We use the pinned obstacles as a means of experimentally-recreating real-world examples of crowd dynamics, such as those observed funneling through gates with some organized, persistent motion following the obstacles (Fig. 1b-d, Visualization 1).

Because our LC skyrmions represent minima in the elastic free energy of our chiral nematic experimental systems, we minimize the Frank-Oseen free energy to computer simulate the **n(r)** of a single skyrmion at electric fields **E** corresponding to different applied voltages U: [9,20,35,36]

$$W = \int \left\{ \frac{K_{11}}{2}(\nabla \cdot \mathbf{n})^2 + \frac{K_{22}}{2}\left[\mathbf{n} \cdot (\nabla \times \mathbf{n}) + \frac{2\pi}{p_0}\right]^2 + \frac{K_{33}}{2}\left[\mathbf{n} \times (\nabla \times \mathbf{n})\right]^2 - \frac{\varepsilon_0 \Delta\varepsilon}{2}(\mathbf{E} \cdot \mathbf{n})^2 \right\} dV \quad (1)$$

In the expression above, the elastic constants $K_{11}$, $K_{22}$, and $K_{33}$ represent the elastic energy costs for splay, twist, and bend deformations of **n(r)**, respectively. $\Delta\varepsilon$ is the dielectric anisotropy and $\varepsilon_0$ is the permittivity of free space. In this study we use materials with negative dielectric anisotropy such that, upon voltage application across the sample thickness, the director **n(r)** rotates to lie perpendicular to the direction of the applied field **E** in the sample mid-plane, thus deforming our skyrmionic structures from their initial axi-symmetric state to an asymmetric one shown in Fig. 1e. Similar asymmetric skyrmions have been demonstrated in both LCs and chiral magnets [9,37,44-47]. Due to their topological stability, the resulting asymmetric skyrmions maintain their topological nature under tunable deformation [9,36,37]. This morphing of the director field with electric field application is not invariant upon reversal of time, which results in a net translational motion upon voltage modulation [36]. Many skyrmions under such energy conversion conditions tend to exhibit long-range interactions and schooling behavior [9]. Vectorized **n(r)** plots (Fig. 1e), midplane director field schematics displaying the **n(r)** orientations in the form of nonpolar field lines (Fig. 1f), and numerically-generated polarizing images (Fig. 1g) are all consistent with experimental images of similar structures under similar conditions (Fig. 1h). These computer simulated polarizing optical microscopy images are generated using a Jones matrix method [36,15] in which the optical properties of the material (Table 1) and the sample thickness are taken as numerical inputs for calculating the resulting polarizing optical microscopy textures that one would expect to see. The material parameters used in these computer simulations presented in Fig. 1 correspond to the experimental values for ZLI-2806 with a CB-15 additive (Table 1). Using two different nematic hosts and both left- and right-handed cholesterics (corresponding to different chiral additives, Table 1) allowed us to assure the broad applicability of our findings to a broad range of chiral nematic material systems.

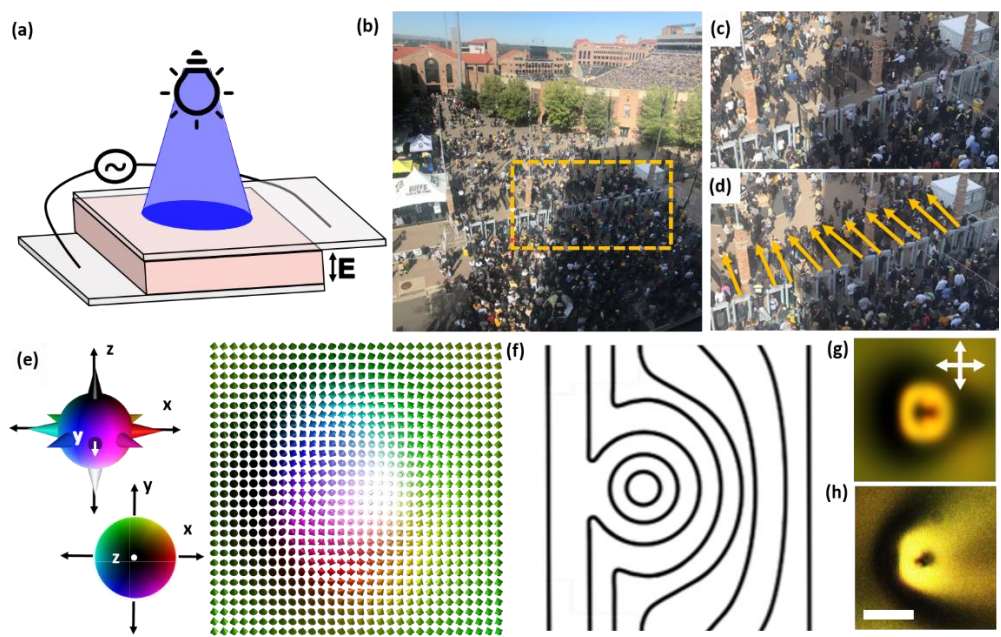

Fig. 1. Experimental setup and motivation. (a) Experimental schematic showing geometry of a liquid crystal sample and blue-light illumination, with direction of the oscillating electric field application between substrates marked as **E**. (b-d) Crowds of people outside of CU Boulder's Folsom Field as they funnel through entry gates. (c) and (d) represent zoom-ins on the area marked with the dotted box in (b). The orange arrows in (d) mark the persistence of straight motion directly following the gates. (e) Computer simulated vectorized director field structure **n(r)** of a skyrmion at U = 3.5 V, colored according to the orientations and points on the $\mathbb{S}^2$ order parameter space (note that the order parameter space for the nonpolar un-vectorized director field is $\mathbb{S}^2/\mathbb{Z}_2$, a sphere with diametrically opposite points identified), as shown on the left. (f) Corresponding visualization of the director field lines in the sample's midplane. (g) Computer-simulated polarizing optical image of the skyrmion shown in (e), with (h) comparison to similar experimental polarizing image at U = 3. 5 V. White double arrows mark the crossed polarizer orientations in an optical microscope; the scale bar is 10 µm.

Experimental images and videos were obtained using an Olympus BX-51 upright microscope with a 10x objective and crossed polarizers inserted above and below the sample. For samples with the QL-76 chiral additive, an additional optical filter (obtained from Edmund Optics) was inserted below the sample to enable red-light optical imaging (blocking the short-wavelength blue and green light) and preclude unintentional reaction of the photo-sensitive dopant to the microscope's white light source. A blue-light projection system comprised of an Epson EMP-730 LC Projector's micro-display was coupled to the microscope setup via dichroic mirrors [13,17] that allowed for projection of focused blue-light patterns into the midplane of the sample with tunable polarization capabilities. Images and videos were captured using a charge-coupled device camera purchased from Point Grey Research, Inc and analyzed using ImageJ open source software (National Institute of Health). Positional data for each skyrmion in motion was extracted using the "wrmTrck" plugin for ImageJ and exported to MATLAB for further analysis.

| Material/Property | MLC-6609 | ZLI-2806 |
|---|---|---|
| $T_{NI}$ (°C) | 91.5 | 87 |
| $\Delta\varepsilon$ | -3.7 | -4.8 |
| $h_{HTP}$ of [additive] ($\mu m^{-1}$) | -60 [QL-76] | +5.9 [CB-15] |
| $K_{11}$ (pN) | 17.2 | 14.9 |
| $K_{22}$ (pN) | 7.5 | 7.9 |
| $K_{33}$ (pN) | 17.9 | 15.4 |
| $n_e$ | 1.5514 | 1.518 |
| $n_o$ | 1.4737 | 1.474 |
| $\Delta n$ | 0.0777 | 0.044 |

**Table 1**. **Material properties of nematic hosts and chiral additives**. Nematic-isotropic transition temperature, $T_{NI}$, dielectric anisotropy, $\Delta\varepsilon$, elastic $K_{11}$, $K_{22}$, $K_{33}$ constants, extraordinary ($n_e$) and ordinary ($n_o$) refractive indices, optical anisotropy, $\Delta n$, and helical twisting powers for each chiral additive reported in brackets for the corresponding nematic hosts, $h_{HTP}$. Positive values of $h_{HTP}$ correspond to right-handed chiral additives and negative values of $h_{HTP}$ correspond to left-handed chiral additives.

### 3. Results and discussion

Through selective optical manipulation using the methods described above for laser tweezer manipulation and blue light photo-patterning, we guide skyrmion collective dynamics by introducing new means of controlling velocity, size, motion directionality, jamming, and reconfigurability of skyrmion assemblies. Due to the fact that these structures are highly energetically favorable and stable within our samples, the skyrmions can be squeezed and morphed, unlike hard colloidal or granular active particles and more like squishy biological cells and organisms that can be crammed together and deformed before regaining their preferred shape. Therefore, we demonstrate highly robust manipulation while the skyrmions are in a non-equilibrium state of motion as they move around obstacles, react to changes in the elastic free-energy landscape due to blue-light exposure, and adapt to these interferences by changing their assemblies and trajectories, as detailed below.

*3.1 Obstacle-induced jamming and crowded skyrmion motion*
First, we utilize the optical laser tweezers to carefully arrange a number of obstacles in the path of skyrmion motion trajectories to investigate the crowding and jamming behavior of active skyrmions. Having high levels of control and tunability over the density and position of obstacles and the number of skyrmions moving through the field of view, we start from relatively high densities of each (Fig. 2, Visualization 2). At motion powered by U = 3.5 V and f = 60 Hz, the skyrmions tend to stick together and form quasi-hexagonally-packed clusters which, in this case, leads to large-scale jamming of mobile skyrmions in-between the obstacles. Consequently, velocity within the skyrmion crowd falls to zero and the jamming state persists over a long period of time on the order of minutes.

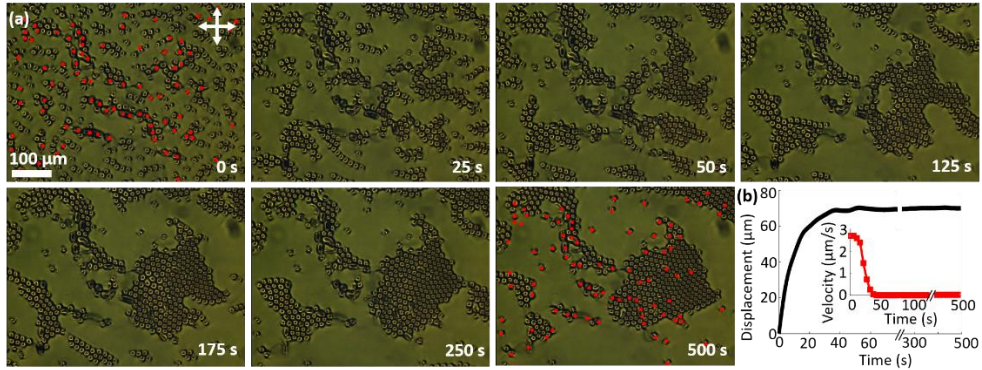

Fig. 2. Jamming skyrmion crowds with optically-induced obstacles. (a) Polarizing optical images corresponding to frames taken from Visualization 2. Motion is powered at by oscillating electric field with voltage U = 3.5 V and oscillation frequency f = 60 Hz. Red points in the first and last frames denote skyrmions selectively-pinned to the confining substrates using optical tweezers. White double arrows mark polarizer orientations in a polarizing optical microscope. (b) Displacement of the skyrmions in the bottom right region as jamming occurs, with corresponding skyrmion velocity shown in the inset.

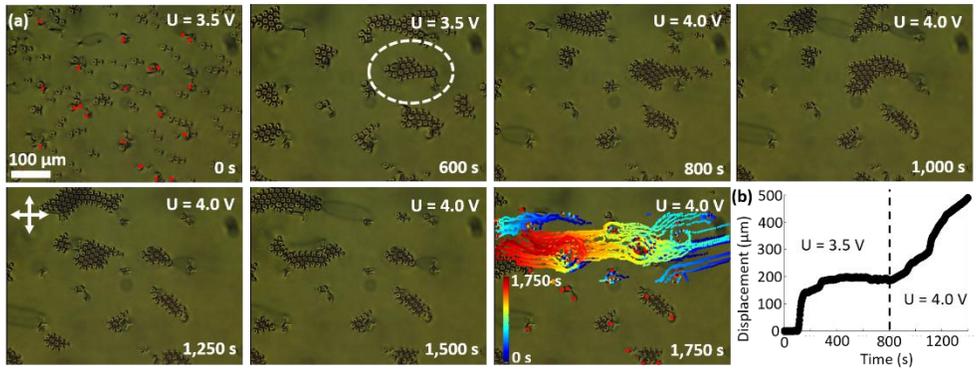

Fig. 3. Using applied voltage to manipulate skyrmion size and overcome jamming. (a) Polarizing images corresponding to frames taken from Visualization 3. The motion is powered by the voltages marked in the top right corners of each frame (with U = 4.0 V corresponding to the smaller skyrmion size) and f = 60 Hz. Red points in the first and last frames denote skyrmions selectively-pinned to the confining substrates using optical tweezers. Skyrmion trajectories are overlaid on the last frame, color-coded according to elapsed time (inset). White double arrows denote the crossed polarizer orientations in an optical microscope. (b) Displacement of the cluster marked with a white dotted circle in part (a) versus time; changing the amplitude of applied voltage from 3.5V to 4.0V at the time marked by the vertical dashed line is used to overcome jamming marked.

We demonstrate, however, that such jamming can be controlled and overcome by tuning skyrmion size via the amplitude of voltage application. We show this using the case of sparser initial skyrmions and obstacles (Fig. 3, Visualization 3), where the jamming still occurs but differently than in the previous case. Once we observe jamming start to occur during the U = 3.5 V motion, we increase the applied field to U = 4 V, which induces more squeezing of the asymmetric skyrmion structures [34] and allows them to compress themselves into closer-packed clusters and overcome the jamming. The clusters at higher voltage, which are comprised of skinnier skyrmions of smaller lateral dimensions, then regain their motion and smoothly traverse through the remaining obstacles.

Instead of being used to induce jamming, the obstacles can also be created in a way that mediates continuous, uninterrupted motion of many skyrmions at once. By organizing the obstacles into channels (Fig. 4, Visualization 4), the skyrmions can be funneled from a dispersed "crowd" into neat single-file lines. Similar to crowds of people passing through security gates (Fig. 1b-d), the skyrmions demonstrate short-range persistence of the directional motion induced by the obstacles (Fig. 4b).

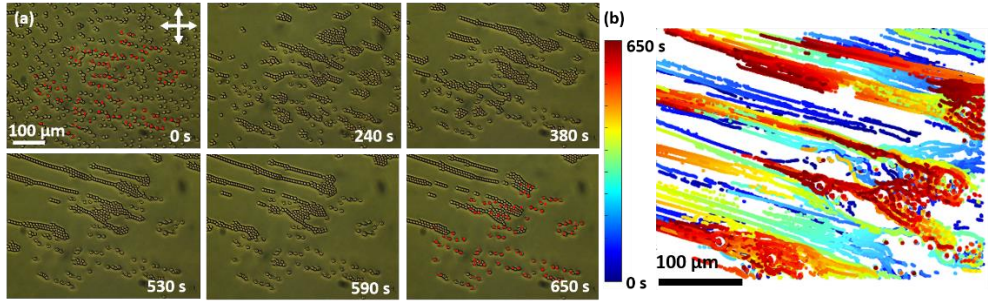

Fig. 4. Re-configuration of skyrmions in motion by pinned obstacles organized to form narrow passages. (a) Polarizing optical images, extracted as frames from the Visualization 4, of individual skyrmions being funneled into long chains by means of selectively-pinned obstacles marked with red points in the first and last frames. The motion is electrically powered at U = 3.5 V and f=60 Hz. White double arrows denote polarizer orientations. (b) Time-coded trajectories of skyrmion motion through the pinned channels, corresponding to the color-coded time scheme shown in the left-side inset.

## *3.2 Photo-induced guiding and deflection of skyrmion motion*

While laser-induced obstacles provide a precise and persisting means of manipulation, they also require hands-on, one-by one creation that is time consuming on the large scale and are not easily modified or moved. We therefore turn to photo-manipulation to add a temporary and adaptable means of controlling skyrmion motion to our toolbox. We characterize the polarization-dependent response of the LC samples with dynamic skyrmions to blue light exposure when the chiral LC is further mixed with the azobenzene-based dopant QL-76. Upon exposure, the illumination light not only increases the helical pitch of the projected area, but the photoresponsive azobenzene-based dopant molecules likewise tend to orient themselves perpendicular to the polarization of the blue light [48], which we can control with a linear polarizing element in our optical microdisplay projection setup [13]. The result of this reorientation is a region within the sample with changing the in-plane tilt directionality of the in-plane **n(r)** in the center of the sample (Fig. 5a,b). This is particularly useful for steering skyrmion motion because, upon voltage modulation, they tend to move perpendicular to the direction of the far-field tilt (Fig. 5c-d), and, thus, along the linear polarization direction.

We demonstrate experimentally, first with a few skyrmions (Fig. 5e,f) and then with many more (Fig. 5g,h), that this in-plane field reorientation induces deflection of motion towards the direction of the polarization. The effect can be understood as resulting from the realignment of the LC director in the cell midplane to orient orthogonally to the linear polarization of illumination light (a common effect for LCs doped with azobenzene-containing dyes [49]), which, in turn, redirects the motion of squirming skyrmions. As skyrmions approach this sample's region of illumination, they slow down from 0.30 µm/s to 0.25 µm/s then, upon overcoming the energetic barrier induced by the blue light, accelerate to 0.36 µm/s within the illumination area. Once outside of the blue-light region, the velocity falls back to the average ~ 0.30 µm/s (Fig. 5e,f). This dynamic behavior can be understood as a combined effect due to both tuning pitch with optical illumination and rotating the cell midplane's director to orient orthogonally to the polarization of the blue illumination light.

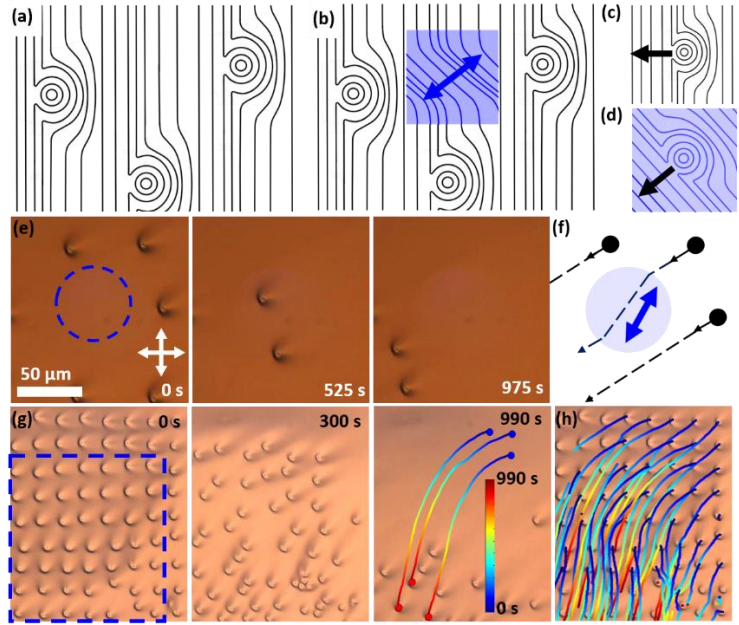

Fig. 5. Dynamic response of skyrmions in motion to patterned linearly polarized light. (a-d) Schematics of the midplane director field showing the director orientation in the midplane of the sample for (a) a region with asymmetric skyrmions before patterned illumination and (b) the same region with skyrmions and a small exposure area (blue); the corresponding change in the midplane director field is shown with blue lines within the area where the midplane director field turns to be perpendicular to the polarization direction of the illumination light (blue double arrow). (c-d) Schematics of the midplane director field for a single skyrmion in motion (c) without illumination and (d) with illumination, where the polarization is the same as in part (b). Black arrows mark the direction of motion. (e) Polarizing optical images of a few skyrmions in motion, with the skyrmion that passes through the illuminated region (marked with dotted circle in the first frame) being deflected (re-directed) by the light. (f) Corresponding schematic illustrating the skyrmion trajectories with dotted lines, illumination area in blue, and the blue-light polarization with a double-headed blue arrow. (g) Polarizing optical images of a lattice of skyrmions in motion, with the illuminated region marked with a dashed blue box. Color-time-coded trajectories in the last frame show the starting (blue dots) and ending positions (red dots) for three skyrmions. The inset in the 3$^{rd}$ frame shows the time-color-coded scale. (h) The initial polarizing image from (g) with all skyrmion trajectories overlaid and colored according the inset in the 3$^{rd}$ frame of (g). Motion in (e-h) is powered by applied oscillating electric field at U = 3.5V with f = 1 kHz (carrier frequency), modulated at 2Hz. White double arrows in (e) mark polarizer orientations.

Our experimental projection setup enables application of various illumination shapes and patterns, which we utilize to show the precise tunability of directional steering. To do this, a pattern was chosen in which a small central channel has no exposure while two semi-circles above and below the channel are exposed to induce in-plane **n(r)** reorientation (Fig. 6, Visualization 5). As the self-assembled chains move towards the left side of the frame in Fig. 6, one chain in particular travels into the channel then "feels" the exposure near the center of the channel, where the reorientation is arguably the strongest, and then rotates to travel upwards in a trajectory parallel to the direction of the linear polarization of illumination light (Fig. 6).

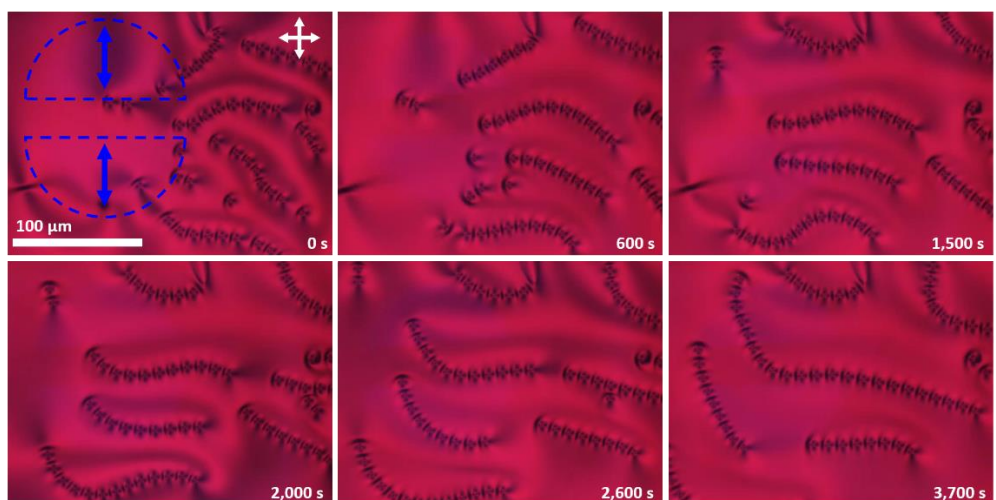

Fig. 6. Directional steering of dynamic skyrmion chains using patterned light illumination. Polarizing optical images, taken as frames from the Visualization 5, of self-assembled chains of skyrmions in motion at U = 3.5 V and f = 1kHz (carrier frequency), modulated at 2Hz. Blue dashed lines in the first frame mark the sample region exposed to patterned light illumination, with blue double arrows marking the direction of the blue light's linear polarization. White double arrows mark the optical microscope's crossed polarizer orientations.

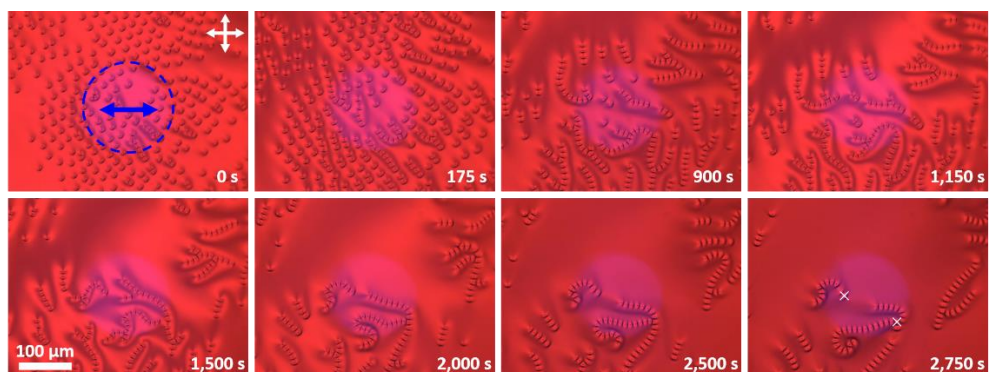

Fig. 7. Reconfigurable motion and self-assembly of skyrmions under light illumination interacting with pinned obstacles. Polarizing images taken as frames from Visualization 6, where motion is powered at U = 3.5 V and f = 1 kHz (carrier frequency) modulated at 2 Hz. The blue dashed circle in the 1$^{st}$ frame denotes the illumination area with a blue double arrow marking the direction of blue-light linear polarization. White double arrows mark the microscopy polarizer orientations. Two pinned skyrmions are marked with white crosses in the last frame.

### 3.3 Reconfiguring skyrmion motion using laser-patterned obstacles and light

We have now developed an experimental system with multifaceted, adjustable properties that can be used to influence skyrmion motion. Next, we investigate how a combination of these semi-permanent and short-term control tools can be used to facilitate new emergent behavior within the collective migration of school-like assemblies. One such example of emergent behavior lies in the dynamic rearrangement of skyrmion chains (Fig. 7, Visualization 6), in this case utilizing both pinned obstacles and the steering mechanism induced by blue-light illumination, with the polarization direction set to be perpendicular to the un-exposed trajectories of motion. In this case, the laser-pinned obstacles act as nucleation sites for the oscillatory motion of chains that elastically attach themselves to the stationary obstacle sites and then swirl and reconfigure their snake-like assemblies, moving in and out of the exposure area and thus individually changing their orientations and motion directions.

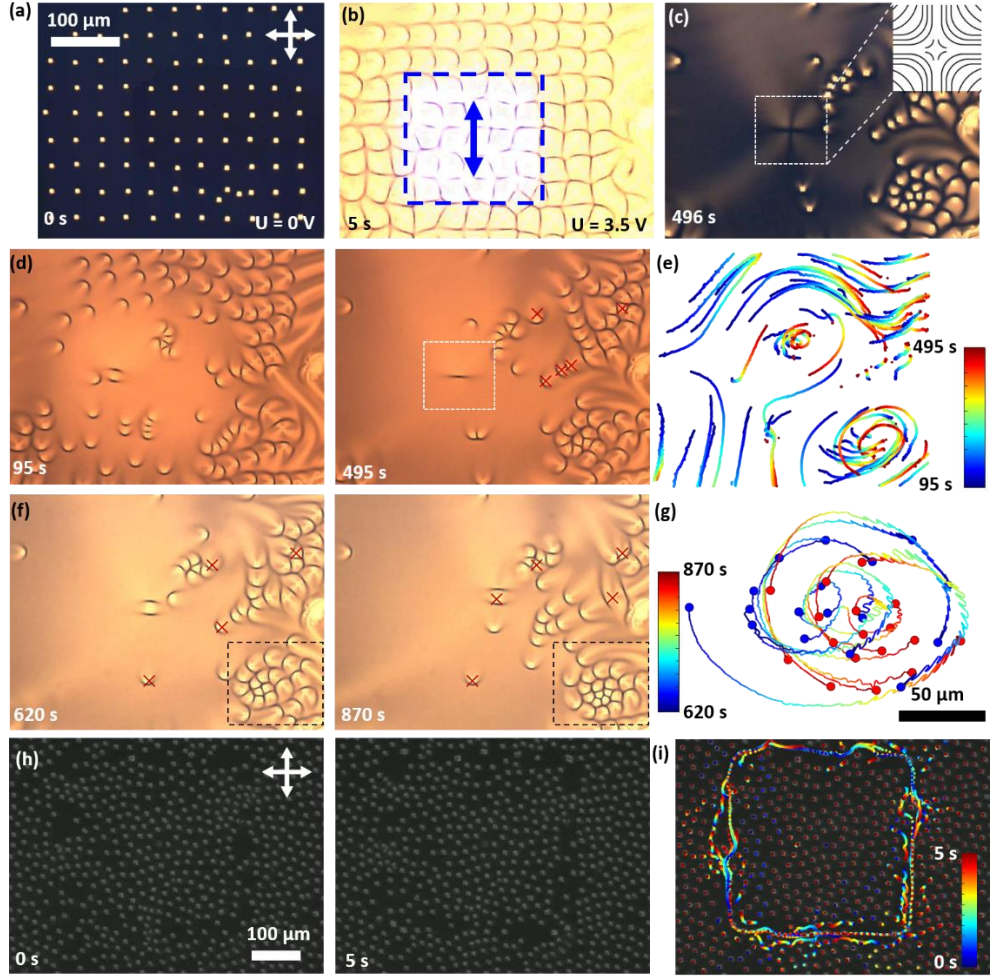

Fig. 8. Complex spiraling dynamics of skyrmions defined by light and oscillating electric fields. (a-c) Polarizing optical images of the skyrmion lattice (a) prior to voltage application at U = 0 V and (b) at the onset of voltage application at U = 3.5 V and f = 1kHz (carrier frequency), modulated at 2 Hz. Patterned blue-light illumination is applied at the onset of motion for 10s within a sample region defined by the blue dashed box, with the blue double arrow denoting polarization of the illumination light. (c) Polarizing optical image obtained during voltage modulation and showing the texture of an unpaired umbilical defect (containing 4 dark and 4 bright brushes), with the corresponding field director profile (inset). (d) Polarizing optical images during complex skyrmion motion, where the dashed white box marks the same umbilic defect as described in part (c), but now observed in a brightfield imaging mode. (e) Trajectories of the skyrmion motion for the entire field of view shown in part (d), which are color-coded with time according to the scheme in the inset. (f) Polarizing optical images obtained during motion, where the dashed black box marks a region in which there is significant swirling motion, also shown as (g) trajectories color-coded with time according to the scheme in the inset. Blue and red dots mark the starting and ending positions for each skyrmion, respectively. Red crosses in the experimental frames in (d,f) mark skyrmions that have been caught, randomly and sometimes temporarily, on surface defects. (h) Polarizing optical images obtained during selective infrared-beam-guided motion (the elapsed time is marked in the bottom-left of the images) and the corresponding (i) trajectories color-coded with time according to the bottom-right inset. White double arrows mark polarizer orientations. The scale bar in (a) refers to images (a-f) and that in the 1st frame of (h) to optical images in (h-i).

Finally, we present a demonstration of emergent dynamic behavior featuring patterned exposure, dynamic structures that become obstacles at random by getting stuck on surface defects (laser-induced pinning sites), and an additional unpaired topological defect called an umbilic (Fig. 8, Visualization 7). This elementary-winding-number umbilic is a nonsingular defect in the in-plane **n(r)** tilt directionality field that is created spontaneously upon voltage application and self-assembles with the twisted structure of the skyrmion core to stabilize the elementary skyrmion in the tilted far-field (Fig. 8c, inset) [37]. In this case, we create the "unpaired" umbilic by exposure to blue light at the transition between the voltage off and on states [17]. Because the unpaired umbilic represents a long-range spatial pattern in the tilt director directionality field, when skyrmions in motion come near the defect, they are deflected and sidetracked according to the direction of the local tilt directionality field, moving orthogonally to it. This likewise has long-range effects on the trajectories of motion, which develop complex swirling pattern that persists over long periods of time (Fig. 8d-g). During this motion, random skyrmions in motion temporarily get stuck on various pinning sites (marked by red crosses within the experimental images in Fig. 8d,f) and, favorably, can free themselves from the pinning sites and persist with their emergent collective movement. This finding demonstrates that the technique for pinning obstacles with laser tweezers can either be used as a permanent or a temporary means of hindering and adjusting trajectories of motion. This observation also provides a number of new experimental knobs to turn for controlling and guiding active topological solitons. In particular, one can envisage optical or other types of patterning of the director field in the midplane that can lead to electrically powered transport of LC skyrmions along well-defined pre-programmed trajectories. Although trajectories of motions of similar skyrmions can be defined by real-time scanning infrared laser beams of optical tweezers with velocity that allows for skyrmions following the laser traps without escaping them (Fig. 8h-i), the demonstrated capability of combining ambient-light-based control of electrically powered skyrmion motions (Fig. 8a-g) can be used over much larger areas and without sophisticated laser tweezer setups. Additionally, it can be combined with the very same laser trap scanning to manipulate skyrmion dynamics in even more versatile ways, for example, to accelerate and re-direct skyrmion motions within localized sample regions (Fig. 8i).

## 4.  Conclusions

In this work, we have established multifaceted means of control over topological solitons in collective emergent motions by manipulating the velocity, size, directionality, jamming and sorting behavior, and through the reconfigurability of skyrmion assemblies. We have accomplished this by utilizing a combining optical manipulations that have never been used in harmony to enhance this active behavior and uncover new means of experimental control. By selectively pinning topological solitons to the surface alignment layer on the glass substrates using a focused optical laser tweezer setup, we create semi-permanent stationary obstacles around which skyrmions can move smoothly or exhibit jamming behavior or movement reminiscent of crowds of people funneling through security gates. A photo-sensitive chiral additive enables precise steering and accurate manipulation of motion for both individual structures and self-assembled groups of skyrmions upon exposure by tuning the polarization of the blue-light illumination. When these techniques are used together, we gain multifaceted control over trajectories of motion and oscillatory-like behavior such as slithering and swirling of skyrmion chains. As these structures are energetically favorable, easily stabilized, and highly robust in optically anisotropic LC materials, like the ones used in displays, there are many exciting opportunities for applications development and new touch-screen technologies based on emergent responses to external fields and selective manipulation of twisted solitons. The experimental sample fabrication techniques used throughout this study are highly reminiscent of those used in the multi-billion-dollar LC display industry, which can potentially stimulate the translation of our findings to the consumer markets. Previous studies have shown the ability to use specialty experimental systems involving micropatterned substrates [50], thickness gradients, and structured beams of light to control skyrmionic structures in equilibrium conditions [28,35,50-52], however our work enables higher levels of dynamic control using low-intensity unstructured light. These advances add to the experimental toolkit available for controlling active behavior of solitons and promises new complex avenues and compelling possibilities for technological uses for LC skyrmions. From the active matter standpoint, where guiding active particles by microfabricated topographic pathways and other means has been a mission of many recent studies [2,53-56], our light-guided active skyrmions can serve as a model system to probe such emergent out-of-equilibrium behavior in new regimes of external control. The dynamic skyrmions in racetrack magnetic memories [25,26] may potentially pin on impurities and defects in crystalline solid films and, therefore, our work on topologically similar LC skyrmions [57] may potentially provide insight into ways of unjamming the moving skyrmions in the contexts of such applications.

**Funding**

This research was supported by the National Science Foundation (NSF) through Grants No. DMR-1810513 (research) and No. DGE-1144083 (graduate research fellowship to H.R.O.S.). The computation facilities utilized in this work were supported by the NSF grants ACI-1532235 and ACI-1532236 (RMACC Summit supercomputer used for the numerical modeling).

**Acknowledgements**

We thank T. Bunning and T. White for providing the QL-76 chiral dopant used throughout this study. We also thank P.J. Ackerman, A. Hess, J-S. B. Tai and Y. Yuan for useful discussions.

**Disclosures**

The authors declare no conflicts of interest.

## Visualization Captions

Visualization 1: Time lapse of crowds funneling through gates at CU Boulder's Folsom Field. The video was taken on an iPhone 7 and the total elapsed time is ~ 10 minutes.

Visualization 2: Jamming skyrmion crowds with optically-induced obstacles. Polarizing optical microscopy video of motion obtained at U = 3.5 V and f = 60 Hz. Obstacles were selectively pinned by optical tweezers, with their exact positions shown in Fig. 2. White double arrows mark polarizer orientations. The video is played at ~30x speed. The total elapsed time is 500 s.

Visualization 3: Using skyrmion size manipulation to overcome jamming. Polarizing optical microscopy video of motion powered by the voltages marked in the top left and f = 60 Hz. Obstacles were selectively pinned by optical tweezers, with their exact positions labeled in Fig. 3. White double arrows mark polarizer orientations. The video is played at ~110x speed. The total elapsed time is 1,750 s.

Visualization 4: Re-configuration of skyrmions in motion by pinned obstacles organized to form narrow passages. Polarizing optical microscopy video of motion powered at U = 3.5 V and f = 60 Hz. Obstacles were selectively pinned by optical tweezers, with their exact positions plotted in Fig. 4 White double arrows denote crossed polarizer orientations. The video is played at ~30x speed. The total elapsed time is 650 s.

Visualization 5: Directional steering of skyrmion chains using patterned and polarized light illumination. Polarizing optical microscopy video of self-assembled chains of skyrmions in motion at U = 3.5 V and f = 1 kHz (carrier frequency) modulated at 2 Hz. Details of patterned illumination are provided in Fig. 6. White double arrows mark the polarizing optical microscope's linear polarizer orientations. The video is played at ~60x speed. The total elapsed time is 3,700s.

Visualization 6: Reconfigurable motion and self-assembly of skyrmions under light illumination interacting with pinned obstacles. Polarizing optical microscopy video of collective motion at U = 3.5 V and f = 1 kHz (carrier frequency) modulated at 2 Hz. Details of patterned illumination are provided in Fig. 7. White double arrows mark the microscope's linear polarizer orientations. The video is played at ~70x speed. The total elapsed time is 2,750 s.

Visualization 7: Complex dynamics of skyrmions following pre-exposure of a skyrmion lattice. Polarizing optical microscopy video, starting from U = 0 V, of collective motion at U = 3.5 V and f = 1 kHz (carrier frequency) modulated at 2 Hz. Voltage is marked in the top left corner of the video. Details of patterned illumination are provided in Fig. 8. White double arrows mark the optical microscope's crossed linear polarizer orientations. The video is played at ~15x speed. The total elapsed time is 870 s.